\documentclass[twocolumn, pra,showpacs]{revtex4}

\usepackage{dcolumn,bm,graphicx}
\usepackage{amsmath}
\usepackage{amsfonts}
\usepackage{amssymb}
\begin{document}
\preprint{UNR 2003--\today }

\title{Marked influence of the nature of chemical bond on
CP-violating signature in molecular ions $\mathrm{HBr}^{+}$ and $\mathrm{HI}^{+}$}
\author{Boris Ravaine}
\author{Sergey G.~Porsev}
\altaffiliation{Petersburg Nuclear Physics Institute, Gatchina,
Leningrad district, 188300, Russia}
\author{Andrei Derevianko}
\email{andrei@unr.edu}
\affiliation {Department of Physics, University of Nevada, Reno, Nevada 89557}

\date{\today}

\begin{abstract}
Heavy polar molecules offer a great sensitivity to the electron Electric Dipole Moment(EDM).
To guide emerging searches for EDMs with molecular ions,
we estimate the EDM-induced energy corrections for hydrogen halide ions
$\mathrm{HBr}^{+}$ and $\mathrm{HI}^{+}$ in their respective ground
$X\,^2\!\Pi_{3/2}$ states.
We find that the energy corrections due to EDM for the two ions differ by an unexpectedly
large factor of
fifteen.
We demonstrate that a major part of this enhancement is due to a dissimilarity
in the nature
of the chemical bond for the two ions: the bond that is nearly of ionic character in $\mathrm{HBr}^{+}$
exhibits predominantly covalent nature in $\mathrm{HI}^{+}$. We conclude that because of this
enhancement the HI$^+$ ion may be a potentially competitive candidate for the EDM search.
\end{abstract}


\maketitle

A non-vanishing permanent electric dipole moment (EDM) of a particle simultaneously
violates two discrete symmetries: parity (P) and time reversal
(T). By the virtue of the CPT theorem, the T-violation would imply
CP-violation~\cite{BigSan00}. Searches for
EDMs of atoms and molecules~\cite{KhrLam97} provide important
constraints on competing extensions to the standard model of
elementary particles. For example, the most stringent
limits on electron EDM come from a table-top experiment with atomic Tl~\cite{RegComSch02}.
As with atoms, the internal energy states of heavy polar molecules
can show evidence of EDMs of the constituents.
Compared to atomic experiments, where application of
strong external E-field is required to enhance sensitivity,
the experiments with polar molecules
rather rely on the inner electric molecular field $\mathcal{E}_\mathrm{int}$
exerted upon the heavier atom.
This field can be several orders
of magnitude larger than the attainable laboratory fields.
This notion, first elucidated by \citet{San67}, has been exploited in  experiments
with YbF~\cite{HudSauTar02} and TlF~\cite{WilRamLar84,SchChoVol87,ChoSanHin89} molecules.
We also mention ongoing experiment with
metastable PbO molecule\cite{KawBayBic04}.

A relatively small laboratory field is
still required in the EDM experiments to polarize the molecule.
Since the E-field would accelerate a charged particle out of
an apparatus, EDM experiments are typically
carried out using neutrals.
It has been recently realized by \citet{StuCor04} that this
limitation may be overcome with ion traps:
electrostatic force exerted upon the ion by the polarizing E-field
can average to zero if the polarizing field rotates rapidly in space,
with the requisite spectroscopy then being performed in a rotating frame of reference.
Moreover, the long coherence times in the trap
would improve statistics compared to traditional beam and gas cell approaches.
Because of this improved statistics,
molecular ions with a relatively weak sensitivity to electron EDM could provide
competitive constrains.
In particular, the hydrogen halide  ions  $\mathrm{HBr}^{+}$ and
$\mathrm{HI}^{+}$ in their lowest rovibrational state of the ground
$X\, ^2\!\Pi_{3/2}$ term are considered as attractive candidates for the proposed
experiment~\cite{StuCor04}.

The goal of this paper is two-fold. Firstly, we provide a guidance
to emerging EDM searches with molecular ions~\cite{StuCor04} by computing EDM-induced
energy corrections. Secondly, we elucidate
the important role of the chemical bond in enhancement of electron EDM
in molecular systems. While both $\mathrm{HBr}^{+}$ and $\mathrm{HI}^{+}$
ions have a similar electronic structure, the chemical bond in
$\mathrm{HBr}^{+}$ is of ionic nature, while for heavier $\mathrm{HI}^{+}$
it is predominantly covalent\cite{ChaHoDal95a,ChaHoDal95b}.
We find that this evolution in the character of the chemical bond
has a marked effect on the EDM-induced energy corrections.
From the experimental point of view, our computed  EDM-induced
energy correction for HBr$^{+}$ is too small to produce
competitive bounds on the electron EDM in experiment~\cite{StuCor04}. By contrast,
the pronounced covalent bond enhancement for the HI$^+$ ion, illuminated here,
makes it a potentially competitive candidate for the electron EDM search.

{\em Molecular structure and EDM-induced corrections.}
The molecular structure of low-lying rotational states of hydrogen halide ions HBr$^+$ and HI$^+$
can be well classified by the Hund's case (a). Relativistic effects split the
ground $X\,^2\Pi$ electronic term  into two components: $^2\Pi_{3/2}$ and
$^2\Pi_{1/2}$, distinguished by $\Omega$, projection of the total electronic angular
momentum along the molecular axis.
$^2\Pi_{3/2}$ is the ground electronic
term and it is considered as a possible candidate for the
EDM experiment.
In the estimates below we will employ the following values of the equilibrium internuclear
separations~\cite{ChaHoDal95a,ChaHoDal95b}: $R_e \approx 1.448~\mathring{A}$ for HBr$^+$
and $R_e \approx 1.632~\mathring{A} $ for HI$^+$. Unless noted otherwise,
atomic units, $\hbar=|e|=m_e\equiv 1$, are used throughout the paper.

In Hund's case (a) the molecular eigenfunctions including the nuclear rotation
can be described as
$\left|\Lambda \Sigma \Omega; J M_J \right\rangle =
\left\vert JM_{J}\Omega \right\rangle \left\vert \Lambda \, \Sigma \, \Omega \right\rangle\,
$, where
$\Lambda$ and $\Sigma$ are projections of the electronic orbital momentum and spin
onto the internuclear axis,
$J$ is the total molecular
momentum (including nuclear rotation) and $M_J$ is the laboratory frame projection of $J$.
The rotational part $\left\vert JM_{J}\Omega \right\rangle$ may be expressed in terms of the Wigner $D$ functions.
While in the lowest-order $\left|\Lambda \Sigma \Omega; J M_J \right\rangle$ and
$\left|-\Lambda -\Sigma -\Omega; J M_J \right\rangle$ states have the same energies,
at the finer level each rotational state of the $^2\Pi_\Omega$ terms splits into so-called
$\Lambda$--doublet~\cite{LefFie86} due to
rotational and spin-orbit perturbations. The eigenstates of the field-free molecular
Hamiltonian (disregarding EDM) are $e/f$ parity states, composed of linear
combinations of the two above states.

An externally applied electric field $\mathcal{E}_{0}$
couples the $e/f$ parity states. For a sufficiently strong E-field
the eigenstates
can be classified by a definite value of $\Omega$, rather than by the $e/f$ parity label.
In this case the correction to the energy due to electron EDM can
be parameterized as~\cite{KozDem02}
$\delta W(J,M,\Omega)= W_{d}\Omega.$
It is defined as an expectation value
\begin{equation}
W_{d}\Omega =
\left\langle \Lambda \Sigma \Omega; J M_J \right|
H_{e} \left|\Lambda \Sigma \Omega; J M_J \right\rangle =
\langle \Lambda \, \Sigma \, \Omega \vert H_{e}
\vert \Lambda \, \Sigma \, \Omega \rangle \, . \label{Eq:Wd}
\end{equation}
Here $H_e=-d_e(\gamma_0-1)\mathbf{\Sigma}\cdot \mathbf{\mathcal{E}} $ is
the pseudo-scalar coupling~\cite{KhrLam97} of
an electron EDM $d_e$ to an electric field $ \mathbf{\mathcal{E}} $
(this internal molecular field is to be
distinguished from the externally applied field).
The expectation value~(\ref{Eq:Wd})  is accumulated in the region of strong
fields, i.e., mainly in the vicinity  of the nucleus
of the heavier halogen atom. A common approximation is that the electric field is produced by a
spherically symmetric charge distribution
$
\mathbf{\mathcal{E}(\mathbf{r})
}\approx\ Z/r^{2} \, \mathbf{\widehat{r}} \, ,
$
where $Z$ is the nuclear charge of the heavier atom, and $\mathbf{r}=0$  coincides with
its center.

{\em Chemical bond.} In the following we make an order-of-magnitude estimate of the EDM
factor $W_{d}$ using a qualitative model of an isolated atomic particle perturbed
by its molecular counterpart.
In this regard it is important to discuss
the nature of the chemical bond in the hydrogen halide HX$^+$ ions. It can be
be described by two limiting cases~\cite{ChaHoDal95b}: ionic (H$^+\!\!:\,$X) and covalent (H $\cdots$ X$^+$) bonds.
In the case of the ionic bond the halogen atom is electrostatically perturbed by a proton.
When the bond is covalent, the halogen atom is singly ionized ($^3P$ state),
while the hydrogen atom is in its ground state.

Although both HBr$^{+}$ and HI$^+$ ions dissociate to the covalent limit,
the chemical bond at intermediate separations
can be better characterized from molecular spectra. In particular,
the hyperfine structure is of significance to our consideration,
because both the EDM coupling and the hyperfine interaction
are sensitive to  behavior of the molecular orbitals near the nuclei.
An analysis of the hyperfine structure in Ref.~\cite{ChaHoDal95a}
indicates that the bond for the HBr$^{+}$ ion can be adequately
described as being of the ionic nature.
As to the HI$^+$ ion, the hyperfine-structure analysis by the same authors~\cite{ChaHoDal95b}
shows that the bond is predominantly of the covalent character.

Below we consider both ionic and covalent bonds. Our semi-qualitative calculations
follow a general scheme similar to those described in Ref.~\cite{KozLab95,KhrLam97}. Firstly, we determine the
effective molecular electric field $\mathcal{E}_\mathrm{int}$ exerted upon
the heavier halogen atom/ion. Then we use the first-order perturbation
theory in the interaction with  $\mathcal{E}_\mathrm{int}$ to determine mixing of the atomic
states of opposite parity. Finally we compute the expectation value
of the EDM-coupling operator using {\em ab initio} relativistic atomic structure codes.

{\em Ionic bond approximation for the $\mathrm{HBr}^+$ ion.}
In the case of the ionic bond the halogen atom is electrostatically perturbed by a proton.
In the spirit of the LCAO\footnote{Linear Combination of Atomic Orbitals} method we
expand the electronic wavefunction in terms of atomic states $\Phi_{i}$  of the
halogen atom
\begin{equation}
\vert \Lambda \, \Sigma \, \Omega \rangle =
\sum_i c_i |\Phi_{i} \rangle \, ,
\end{equation}
where the total angular momentum $J_{e,i}$ of the atomic state  $\Phi_{i}$
and its projection on the molecular axis $M_{e,i}$ are constrained to
$J_{e,i} \ge |\Omega|$ and  $M_{e,i}=\Omega$.
We determine the expansion coefficients $c_i$ using the
first order perturbation theory in the interaction $V$ due to electrostatic
field exerted upon the halogen atom by the proton
\begin{equation}
\vert \Lambda \, \Sigma \, \Omega \rangle \approx
|\Phi_{0} \rangle + \sum_{i \neq 0} |\Phi_{i} \rangle
\frac{\langle \Phi_{i}|V|\Phi_{0} \rangle}{E_{0}-E_{i}}
\label{Eq:pert} \, ,
\end{equation}
where $\Phi_{0}$ is the ground atomic state of the
proper symmetry and $E_{i}$ are the energies of atomic
states.

Keeping only the leading dipole term in the multipole expansion of the
interaction of atomic electrons of the halogen atom with the proton, the
perturbation $V \approx -\mathbf{D \cdot \mathcal{E}}_\mathrm{int}$,
$\mathbf{\mathcal{E}}_\mathrm{int}$ being the electric field of the
proton at the position of the atom and $\mathbf{D}$ the
atomic electric dipole operator. It is this strong electric field that produces enhancement of the
electron EDM in molecular ions.



Finally, the EDM-induced energy correction is
\begin{equation}
 W_{d} \Omega=\frac{2}{R_{e}^{2}} \sum_{i \neq 0} \frac{\langle
\Phi_{0}|H_{e}|\Phi_{i} \rangle \langle\Phi_{i}|D_{z}|\Phi_{0}
\rangle}{E_{0}-E_{i}} \, . \label{Eq:sum}
\end{equation}
In the following we will use a shorthand notation
\begin{equation}
T=H_{e} \, (E_0-H_a)^{-1} \, D_{z} \, , \label{Eq:Toperator}
\end{equation}
with $H_a$
being  the atomic Hamiltonian so that
\begin{equation}
 W_{d} \Omega=\frac{2}{R_{e}^{2}} \,
 \langle \Phi_{0}|T|\Phi_{0} \rangle \, .
 \label{Eq:sumShortIonic}
\end{equation}

It is worth noting that all the quantities (except for empirical $R_e$) in the
Eq.~(\ref{Eq:sum}) are atomic ones and  we employ atomic-structure
methods to evaluate this sum. First we employ Dirac-Hartree-Fock (DHF)
approximation and then more elaborate configuration-interaction (CI) method.
All calculations carried out here are {\em ab initio} relativistic.

The halogen atoms Br and I are open-shell systems with one hole in the
outer $np_{3/2}$-shell, $n=4$ for bromine and $n=5$ for iodine. In the DHF
approximation the atomic orbitals $|i\rangle$ satisfy the eigenvalue equation
$h_\mathrm{DHF} |i\rangle = \varepsilon_{i} |i\rangle$, where
the Dirac Hamiltonian $h_\mathrm{DHF}$ includes an interaction with the  field of the nucleus
and the self-consistent field of the electrons.
In the DHF approximation we obtain
\begin{equation}
W_{d} \Omega =
\frac{2}{R_{e}^{2}} \sum_{i}
\frac{\langle g |h_{e}| i\rangle \langle i | d_{z}|g\rangle}
     {\varepsilon_{i}-\varepsilon_{g}} \label{Eq:DHFSumShort} \, ,
\end{equation}
where $g$ denotes the $np_{3/2}$ hole state and
the summation over $i$ extends over a complete set of orbitals, including both
core and virtual orbitals.

Numerically we carried out the summation using
the B-spline pseudo-spectrum technique\cite{JohBluSap88}. The pseudo-spectrum was generated
using the DHF potential of the ground $^2\!P_{3/2}$ atomic state.
In a typical calculation we used a set of basis functions expanded over 100 B-splines,
which provided numerical accuracy sufficient for the goals of this paper.
Among other technical details it is worth mentioning that while integrating
the radial Dirac equation, we used
the potential produced by a nucleus of the finite size.

To investigate a potentially large correlation effects beyond the DHF approximation,
we have also carried out configuration-interaction (CI) calculations for Br
within the active space of
seven 4$s^2\, 4p^5$ valence electrons. In this method,
the many-electron wave  functions were obtained as linear combinations of determinants
composed from as single and double  excitations of the valence electrons from
the active space.
Finally, following Dalgarno-Lewis-Steinheimer
method \cite{DalLew55}, we carried out the summation over intermediate
states in Eq.~(\ref{Eq:sumShortCovalent}) by solving
inhomogeneous many-body Dirac equation inherent to the method
and computed the sum~(\ref{Eq:sum}). More details will be
provided elsewhere.

The resulting DHF value of the EDM-induced energy correction (hole in the $4p_{3/2}$ shell )
\begin{equation}
W_{d} \Omega [ \mathrm{HBr^{+}, ionic, DHF, X \, ^2\!\Pi_{3/2}} ]  = -1.5 \times 10^{-2} d_{e} \, ,
\end{equation}
A similar DHF calculation assuming a hole in the $4p_{1/2}$ shell
leads to a 100-fold increase in the value of the EDM correction
\begin{equation}
W_{d} \Omega [ \mathrm{HBr^{+}, ionic, DHF, X \, ^2\!\Pi_{1/2}} ]  = 1.6 d_{e} \, .
\end{equation}
A large difference in the values of the $W_d \Omega$ parameter for the two cases
can be explained as follows.
The EDM-coupling operator $H_e$ is
a pseudo-scalar: it does not change the total angular momentum of a state,
but flips its parity. For example, if the hole state $g$ has $p_{3/2}$ angular
character, then the intermediate states in Eq.(\ref{Eq:DHFSumShort})
are $d_{3/2}$ orbitals. Similarly, the $p_{1/2}$ hole state requires $s_{1/2}$
intermediate states. It is well known~\cite{KhrLam97}, that since the states
of lower orbital momentum have a larger probability
to be found close to the nucleus, this selection rule has a profound effect on the
order of magnitude of the EDM factor $W_d$.

One may argue that an enhancement of the
EDM factor for the $X ^2\! \Pi_{3/2}$ state may arise due to particle-hole
excitations, when $s_{1/2}$ ($p_{1/2}$)
electron is excited from the core to the $p_{1/2}$ ($s_{1/2}$) orbital.
It is easy to demonstrate in
the DHF approximation,
that while
the individual contributions from such excitations are certainly
large, their sum vanishes. It is the reason why the closed-shell
systems are  largely insensitive to the electron EDM~\cite{KhrLam97},
i.e the EDM-induced energy correction arises only due to an unpaired electron.
Correlations (many-body effects beyond DHF) may
potentially spoil the presented argument and we have carried out the correlated CI
calculations.
The result,
\begin{equation}
W_{d} \Omega [ \mathrm{HBr^{+}, ionic, CI, X \, ^2\!\Pi_{3/2}} ]
 = -2.6 \times 10^{-2} d_{e}\, ,
\end{equation}
is of the same order as the DHF value.

As a reference, here we also present the DHF value for the HI$^+$ ion in the ionic bond approximation
\begin{equation}
W_{d} \Omega [ \mathrm{HI^{+}, ionic, DHF, X \, ^2\!\Pi_{3/2}} ]
= -7.0 \times 10^{-2} d_{e} \, .
\end{equation}

{\em Covalent bond approximation for $\mathrm{HI^{+}}$.}
From the preceding discussion it is clear that a participation of the unpaired
$p_{1/2}$  or $s_{1/2}$ orbital in the
ground-state configuration of the heavier molecular constituent
is important for gaining  large  $W_d$ parameter.
Qualitatively we can  hope that such an enhancement  for
the $X^2\!\Pi_{3/2}$ component  may arise when the chemical bond acquires
covalent character (case of HI$^{+}$ ion). Indeed,
in the covalent bond approximation, the halogen atom becomes singly ionized
its ground state being $^3\!P$. The ground state has two $p$ holes in
the outer shell, so that the corresponding
relativistic many-body states are composed
from linear combination of $p^{-1}_{1/2} \, p^{-1}_{1/2}$,
$p^{-1}_{3/2} \, p^{-1}_{1/2}$ and $p^{-1}_{3/2} \, p^{-1}_{3/2}$ single-
electron configurations (the superscript $-1$ designates a hole state).
Therefore the unpaired $j=1/2$ orbital becomes involved in the calculations,
and indeed, as shown below, this leads to
a significantly larger EDM-induced energy correction for the $X^2\!\Pi_{3/2}$ term.

The HI$^{+}$ ion may be pictured as the iodine ion I$^{+}$ in the $^{3}P$ state perturbed by the
neutral hydrogen atom in its ground state.
First let us derive the internal
electric field $\mathcal{E}_\mathrm{int}$ and the associated mixing of opposite parity
states of the iodine ion. Qualitatively, the  field of I$^{+}$
induces a dipole moment of the hydrogen atom $|D_{H}|=\alpha_{0}/R_e^2$, where
$\alpha_{0}=9/2$ is the polarizability of the hydrogen ground state. In
turn, the induced dipole moment exerts a field at the position of the iodine ion
$\mathcal{E}_{int}= 2\alpha_{0}/R_{e}^{5}\hat{z}$.
Thus the iodine ion is perturbed by
$
V \approx - 2\alpha_{0}D_{z} / R_{e}^{5}  \, ,
$
where $D$ is the atomic dipole moment operator for I$^{+}$.

Again we limit our consideration to a qualitative estimate and
use the first-order perturbation theory in the molecular field, so that
the EDM-induced energy correction is
\begin{eqnarray}
W_{d}\Omega=-\frac{4\alpha_{0}}{R_{e}^{5}} \,
\langle \Phi_{0}|T|\Phi_{0} \rangle \, ,
 \label{Eq:sumShortCovalent}
\end{eqnarray}
where the operator $T$ is given by Eq.~(\ref{Eq:Toperator}),
except now all the participating operators in that expression
are to be understood as being for the iodine ion and $\Phi_{0}$ is its properly
symmetrized ground state. The above expression differs from the analogous formula~(\ref{Eq:sumShortIonic})  for the ionic bond
by a prefactor characterizing the internal molecular field $\mathcal{E}_\mathrm{int}$ acting
upon the halogen atom/ion.
Compared to the ionic bond, this perturbing field becomes 70\% weaker.

In order to carry out the calculations with the relativistic operator $H_e$,
we express the unperturbed non-relativistic molecular wavefunction in terms of
the
relativistic wave-functions of the iodine ion, $\left\vert ^{3} P_{J},M_{I^{+}}\right\rangle$,
 and the hydrogen, $\left\vert 1s_{1/2},M_\mathrm{H} \right\rangle$,
\begin{eqnarray}
& &\left\vert ^{2}\Pi_{3/2}\right\rangle _{\mathrm{covalent}}^{(0)}  = \sqrt{\frac
{2}{3}}\left\vert ^{3}P_{2},2\right\rangle \left\vert 1s_{1/2},-\frac{1}{2}\right\rangle +
 \\
& & -\sqrt{\frac{1}{6}} \left(
\left\vert ^{3} P_{2},1\right\rangle \left\vert
1s_{1/2},\frac{1}{2}\right\rangle
+ \left\vert ^{3}P_{1},1\right\rangle \left\vert
1s_{1/2},\frac{1}{2}\right\rangle \right) \, . \nonumber
\label{Eq:twocenter}
\end{eqnarray}

Since the expectation value of the EDM coupling operator $H_e$  is
accumulated close to the nucleus of the heavy iodine ion, $H_e$ is essentially
a one-center operator and a generalization of
Eq.(\ref{Eq:sumShortCovalent}) for the two-center wavefunction (\ref{Eq:twocenter})
reads
\begin{eqnarray}
\lefteqn{ \left( -\frac{4\alpha_{0}}{R_{e}^{5}} \right)^{-1}
\langle^{2}\Pi_{3/2}|H_{e}|^{2}\Pi_{3/2}\rangle_{\mathrm{covalent}}
=}\\ \nonumber
& &\frac{3}{2}\left\langle ^{3}P_{2},1\right\vert T%
\left\vert ^{3}P_{2},1\right\rangle +
 \frac{1}{6}\left\langle ^{3}P_{1},1\right\vert T%
\left\vert ^{3}P_{1},1\right\rangle +\\ \nonumber
& & \frac{1}{6}\left\langle ^{3}P_{2},1\right\vert T%
\left\vert ^{3}P_{1},1\right\rangle +
 \frac{1}{6}\left\langle ^{3}P_{1},1\right\vert T%
\left\vert ^{3}P_{2},1\right\rangle \, ,
\label{Eq:WdCovalentLong}
\end{eqnarray}
We calculated the values of matrix elements for iodine ion
within the  CI
approach similar to the one described above for Br.
The computed values are
$\left\langle ^{3}\!P_{2},1\right\vert T \left\vert ^{3}%
P_{2},1\right\rangle =  6.4 \, d_{e}$,
$ \left\langle ^{3}P_{1},1\right\vert T \left\vert ^{3}%
P_{1},1\right\rangle = -13.4\, d_{e}$,
$ \left\langle ^{3}P_{1},1\right\vert T \left\vert ^{3}
P_{2},1\right\rangle  = -11.2\, d_{e}$, and
$ \left\langle ^{3}P_{2},1\right\vert T \left\vert ^{3}%
P_{1},1\right\rangle  =  2.5\, d_{e}$.
Finally,
\begin{equation}
W_{d} \Omega [ \mathrm{HI^{+}, covalent,CI, X\,^2\!\Pi_{3/2}} ]  = -0.4 d_{e}\, .
\end{equation}
We notice a sizable enhancement compared to the value of $-7 \times 10^{-2} d_{e}$
obtained in the ionic bond approximation.

{\em Conclusions.} First of all the EDM-induced energy correction for the $X^2\!\Pi_{3/2}$ state
of HI$^+$ is about 15 times larger than for HBr$^+$. A lesser part of this enhancement
comes from the well-known $Z^3$ scaling of CP-violating matrix elements~\cite{KhrLam97}, when
bromine ($Z=35$) is replaced by the heavier iodine ($Z=53$). A more substantial factor,
illuminated in this work, is the evolution in the nature of the chemical bond.
To reiterate, the CP-violating matrix elements are much
larger for the $p_{1/2}$ states than for $p_{3/2}$, due to the fact that
the values of the
relevant matrix elements are accumulated close to the nucleus.
In the ionic bond case of HBr$^+$, the EDM correction arises from an unpaired $p_{3/2}$
hole state in the outer shell and  CP-violating effects are suppressed.
By contrast, the covalent bond of HI$^+$ in addition opens the $p_{1/2}$ shell,
leading to a marked enhancement.

Typical values \cite{Koz97,KozDem02} of the EDM-induced energy corrections  for heavy
{\em neutral} polar molecules PbO and YbF
are on the order of $10 d_{e}$ atomic units. Our computed
value for HI$^+$ is an order of magnitude smaller. Yet, when compared
with the conventional beam and gas-cell experiments, the proposed trapping
experiment~\cite{StuCor04}  has a better statistical sensitivity
so that molecular ions with smaller enhancement parameters, such as HI$^+$,  may suffice.
By contrast, the EDM correction for HBr$^+$ is too small to be of experimental interest.
As shown here, it is the  covalent bond of HI$^+$ that makes this ion a potentially competitive
candidate for the emerging searches for EDMs with molecular ions.

We would like to thank E. Cornell for discussions and for suggesting this problem
and M. Kozlov for comments on the manuscript.
This work was supported in part by the National Science Foundation, by the
NIST precision measurement grant, and
by the Russian Foundation for Basic Research under Grant No.\ 04-02-16345-a.


\begin{thebibliography}{18}
\expandafter\ifx\csname natexlab\endcsname\relax\def\natexlab#1{#1}\fi
\expandafter\ifx\csname bibnamefont\endcsname\relax
  \def\bibnamefont#1{#1}\fi
\expandafter\ifx\csname bibfnamefont\endcsname\relax
  \def\bibfnamefont#1{#1}\fi
\expandafter\ifx\csname citenamefont\endcsname\relax
  \def\citenamefont#1{#1}\fi
\expandafter\ifx\csname url\endcsname\relax
  \def\url#1{\texttt{#1}}\fi
\expandafter\ifx\csname urlprefix\endcsname\relax\def\urlprefix{URL }\fi
\providecommand{\bibinfo}[2]{#2}
\providecommand{\eprint}[2][]{\url{#2}}

\bibitem[{\citenamefont{Bigi and Sanda}(2000)}]{BigSan00}
\bibinfo{author}{\bibfnamefont{I.~I.} \bibnamefont{Bigi}} \bibnamefont{and}
  \bibinfo{author}{\bibfnamefont{A.~I.} \bibnamefont{Sanda}},
  \emph{\bibinfo{title}{CP Violation}} (\bibinfo{publisher}{Cambridge
  University Press}, \bibinfo{address}{Cambridge}, \bibinfo{year}{2000}).

\bibitem[{\citenamefont{Khriplovich and Lamoreaux}(1997)}]{KhrLam97}
\bibinfo{author}{\bibfnamefont{I.~B.} \bibnamefont{Khriplovich}}
  \bibnamefont{and} \bibinfo{author}{\bibfnamefont{S.~K.}
  \bibnamefont{Lamoreaux}}, \emph{\bibinfo{title}{CP violation without
  strangeness. Electric dipole moments of particles, atoms, and molecules.}}
  (\bibinfo{publisher}{Springer}, \bibinfo{address}{Berlin},
  \bibinfo{year}{1997}).

\bibitem[{\citenamefont{Regan et~al.}(2002)\citenamefont{Regan, Commins,
  Schmidt, and DeMille}}]{RegComSch02}
\bibinfo{author}{\bibfnamefont{B.~C.} \bibnamefont{Regan}},
  \bibinfo{author}{\bibfnamefont{E.~D.} \bibnamefont{Commins}},
  \bibinfo{author}{\bibfnamefont{C.~J.} \bibnamefont{Schmidt}},
  \bibnamefont{and} \bibinfo{author}{\bibfnamefont{D.}~\bibnamefont{DeMille}},
  \bibinfo{journal}{Phys. Rev. Lett.} \textbf{\bibinfo{volume}{88}},
  \bibinfo{pages}{071805} (\bibinfo{year}{2002}).

\bibitem[{\citenamefont{Sandars}(1967)}]{San67}
\bibinfo{author}{\bibfnamefont{P.~G.~H.} \bibnamefont{Sandars}},
  \bibinfo{journal}{Phys. Rev. Lett.} \textbf{\bibinfo{volume}{19}},
  \bibinfo{pages}{1396} (\bibinfo{year}{1967}).

\bibitem[{\citenamefont{Hudson et~al.}(2002)\citenamefont{Hudson, Sauer,
  Tarbutt, and Hinds}}]{HudSauTar02}
\bibinfo{author}{\bibfnamefont{J.~J.} \bibnamefont{Hudson}},
  \bibinfo{author}{\bibfnamefont{B.~E.} \bibnamefont{Sauer}},
  \bibinfo{author}{\bibfnamefont{M.~R.} \bibnamefont{Tarbutt}},
  \bibnamefont{and} \bibinfo{author}{\bibfnamefont{E.~A.} \bibnamefont{Hinds}},
  \bibinfo{journal}{Phys. Rev. Lett.} \textbf{\bibinfo{volume}{89}},
  \bibinfo{pages}{023003} (\bibinfo{year}{2002}).

\bibitem[{\citenamefont{Wilkening et~al.}(1984)\citenamefont{Wilkening, Ramsey,
  and Larson}}]{WilRamLar84}
\bibinfo{author}{\bibfnamefont{D.}~\bibnamefont{Wilkening}},
  \bibinfo{author}{\bibfnamefont{N.}~\bibnamefont{Ramsey}}, \bibnamefont{and}
  \bibinfo{author}{\bibfnamefont{D.}~\bibnamefont{Larson}},
  \bibinfo{journal}{Phys. Rev. A} \textbf{\bibinfo{volume}{29}},
  \bibinfo{pages}{425} (\bibinfo{year}{1984}).

\bibitem[{\citenamefont{{D. Schropp, Jr.} et~al.}(1987)\citenamefont{{D.
  Schropp, Jr.}, Cho, Vold, and Hinds}}]{SchChoVol87}
\bibinfo{author}{\bibnamefont{{D. Schropp, Jr.}}},
  \bibinfo{author}{\bibfnamefont{D.}~\bibnamefont{Cho}},
  \bibinfo{author}{\bibfnamefont{T.}~\bibnamefont{Vold}}, \bibnamefont{and}
  \bibinfo{author}{\bibfnamefont{E.}~\bibnamefont{Hinds}},
  \bibinfo{journal}{Phys. Rev. Lett.} \textbf{\bibinfo{volume}{59}},
  \bibinfo{pages}{991} (\bibinfo{year}{1987}).

\bibitem[{\citenamefont{Cho et~al.}(1989)\citenamefont{Cho, Sangster, and
  Hinds}}]{ChoSanHin89}
\bibinfo{author}{\bibfnamefont{D.}~\bibnamefont{Cho}},
  \bibinfo{author}{\bibfnamefont{K.}~\bibnamefont{Sangster}}, \bibnamefont{and}
  \bibinfo{author}{\bibfnamefont{E.}~\bibnamefont{Hinds}},
  \bibinfo{journal}{Phys. Rev. Lett.} \textbf{\bibinfo{volume}{63}},
  \bibinfo{pages}{2559} (\bibinfo{year}{1989}).

\bibitem[{\citenamefont{Kawall et~al.}(2004)\citenamefont{Kawall, Bay, Bickman,
  Jiang, and DeMille}}]{KawBayBic04}
\bibinfo{author}{\bibfnamefont{D.}~\bibnamefont{Kawall}},
  \bibinfo{author}{\bibfnamefont{F.}~\bibnamefont{Bay}},
  \bibinfo{author}{\bibfnamefont{S.}~\bibnamefont{Bickman}},
  \bibinfo{author}{\bibfnamefont{Y.}~\bibnamefont{Jiang}}, \bibnamefont{and}
  \bibinfo{author}{\bibfnamefont{D.}~\bibnamefont{DeMille}},
  \bibinfo{journal}{Phys. Rev. Lett.} \textbf{\bibinfo{volume}{92}},
  \bibinfo{eid}{133007} (\bibinfo{year}{2004}).

\bibitem[{\citenamefont{Stutz and Cornell}(2004)}]{StuCor04}
\bibinfo{author}{\bibfnamefont{R.}~\bibnamefont{Stutz}} \bibnamefont{and}
  \bibinfo{author}{\bibfnamefont{E.}~\bibnamefont{Cornell}},
  \bibinfo{journal}{Bull. Amer. Phys. Soc.} \textbf{\bibinfo{volume}{49}},
  \bibinfo{pages}{76} (\bibinfo{year}{2004}).

\bibitem[{\citenamefont{Chanda et~al.}(1995{\natexlab{a}})\citenamefont{Chanda,
  Ho, Dalby, and Ozier}}]{ChaHoDal95a}
\bibinfo{author}{\bibfnamefont{A.}~\bibnamefont{Chanda}},
  \bibinfo{author}{\bibfnamefont{W.~C.} \bibnamefont{Ho}},
  \bibinfo{author}{\bibfnamefont{F.~W.} \bibnamefont{Dalby}}, \bibnamefont{and}
  \bibinfo{author}{\bibfnamefont{I.}~\bibnamefont{Ozier}}, \bibinfo{journal}{J.
  Mol. Spect.} \textbf{\bibinfo{volume}{169}}, \bibinfo{pages}{108}
  (\bibinfo{year}{1995}{\natexlab{a}}).

\bibitem[{\citenamefont{Chanda et~al.}(1995{\natexlab{b}})\citenamefont{Chanda,
  Ho, Dalby, and Ozier}}]{ChaHoDal95b}
\bibinfo{author}{\bibfnamefont{A.}~\bibnamefont{Chanda}},
  \bibinfo{author}{\bibfnamefont{W.~C.} \bibnamefont{Ho}},
  \bibinfo{author}{\bibfnamefont{F.~W.} \bibnamefont{Dalby}}, \bibnamefont{and}
  \bibinfo{author}{\bibfnamefont{I.}~\bibnamefont{Ozier}}, \bibinfo{journal}{J.
  Chem. Phys.} \textbf{\bibinfo{volume}{102}}, \bibinfo{pages}{8725}
  (\bibinfo{year}{1995}{\natexlab{b}}).

\bibitem[{\citenamefont{Lefebvre-Brion and Field}(1986)}]{LefFie86}
\bibinfo{author}{\bibfnamefont{H.}~\bibnamefont{Lefebvre-Brion}}
  \bibnamefont{and} \bibinfo{author}{\bibfnamefont{R.~W.} \bibnamefont{Field}},
  \emph{\bibinfo{title}{Perturbations in the spectra of diatomic molecules}}
  (\bibinfo{publisher}{Academic Press}, \bibinfo{address}{Orlando},
  \bibinfo{year}{1986}).

\bibitem[{\citenamefont{Kozlov and DeMille}(2002)}]{KozDem02}
\bibinfo{author}{\bibfnamefont{M.~G.} \bibnamefont{Kozlov}} \bibnamefont{and}
  \bibinfo{author}{\bibfnamefont{D.}~\bibnamefont{DeMille}},
  \bibinfo{journal}{Phys. Rev. Lett.} \textbf{\bibinfo{volume}{89}},
  \bibinfo{eid}{133001} (\bibinfo{year}{2002}).

\bibitem[{\citenamefont{Kozlov and Labzowsky}(1995)}]{KozLab95}
\bibinfo{author}{\bibfnamefont{M.~G.} \bibnamefont{Kozlov}} \bibnamefont{and}
  \bibinfo{author}{\bibfnamefont{L.~N.} \bibnamefont{Labzowsky}},
  \bibinfo{journal}{J. Phys. B} \textbf{\bibinfo{volume}{28}},
  \bibinfo{pages}{1933} (\bibinfo{year}{1995}).

\bibitem[{\citenamefont{Johnson et~al.}(1988)\citenamefont{Johnson, Blundell,
  and Sapirstein}}]{JohBluSap88}
\bibinfo{author}{\bibfnamefont{W.~R.} \bibnamefont{Johnson}},
  \bibinfo{author}{\bibfnamefont{S.~A.} \bibnamefont{Blundell}},
  \bibnamefont{and}
  \bibinfo{author}{\bibfnamefont{J.}~\bibnamefont{Sapirstein}},
  \bibinfo{journal}{Phys.\ Rev.\ A} \textbf{\bibinfo{volume}{37}},
  \bibinfo{pages}{307} (\bibinfo{year}{1988}).

\bibitem[{\citenamefont{Dalgarno and Lewis}(1955)}]{DalLew55}
\bibinfo{author}{\bibfnamefont{A.}~\bibnamefont{Dalgarno}} \bibnamefont{and}
  \bibinfo{author}{\bibfnamefont{J.~T.} \bibnamefont{Lewis}},
  \bibinfo{journal}{Proc. Roy. Soc.} \textbf{\bibinfo{volume}{223}},
  \bibinfo{pages}{70} (\bibinfo{year}{1955}).

\bibitem[{\citenamefont{Kozlov}(1997)}]{Koz97}
\bibinfo{author}{\bibfnamefont{M.~G.} \bibnamefont{Kozlov}},
  \bibinfo{journal}{J. Phys. B} \textbf{\bibinfo{volume}{30}},
  \bibinfo{pages}{L607} (\bibinfo{year}{1997}).

\end{thebibliography}
\end{document}